\documentclass{PoS}
\newcommand{\MC}{\multicolumn}
\newcounter{tab}
\title{Revealing evolved massive stars with {\it Spitzer}, {\it WISE} and SALT}

\ShortTitle{Revealing evolved massive stars with {\it Spitzer}, {\it WISE} and SALT}

\author{\speaker{A. Kniazev}\thanks{This paper uses observations made with the Southern African Large Telescope (SALT).} \\
        South African Astronomical Observatory, PO Box 9, 7935, South Africa\\
        Southern African Large Telescope Foundation, PO Box 9, 7935, PO Box 9, 7935, South Africa\\
        E-mail: \email{akniazev@saao.ac.za}}

\author{V. Gvaramadze\\
        Sternberg Astronomical Institute, Lomonosov Moscow State University, Moscow 119992, Russia\\
        Space Research Institute, Russian Academy of Sciences, Profsoyuznaya 84/32, 117997 Moscow, Russia \\
        E-mail: \email{vgvaram@mx.iki.rssi.ru}}

\abstract{We present the results of optical spectroscopic
observations of 54 candidate evolved massive stars revealed
through the detection of mid-infrared nebulae of various shapes surrounding
them with the {\it Spitzer Space Telescope} and {\it Wide-field
Infrared Survey Explorer}. These observations, carried out with
the Southern African Large Telescope (SALT) in 2010--2015, led to
the discovery of about two dozens emission-line stars, of which
15 stars we classify as candidate luminous blue variables (cLBVs).
Spectroscopic and photometric monitoring revealed significant
changes in the spectra and brightness of four newly identified
cLBVs, meaning that they are new members of the class of bona fide
LBVs. We present an updated list of the Galactic bona fide LBVs.
Currently, this list contains eighteen stars, of which more than
70 per cent are associated with circumstellar nebulae. We also
discovered a very rare [WN] star -- the central star of the
planetary nebula Abell\,48, and a WN3 star in a close, eccentric
binary system with an O6\,V star in the Large Magellanic Cloud --
the first-ever extragalactic massive star identified via detection
of a circular shell around it. Most of the remaining targets are
tentatively classified as OB, A and M stars.}

\FullConference{SALT Science Conference 2015 -SSC2015-\\
        1-5 June, 2015\\
        Stellenbosch Institute of Advanced Study, South Africa}

\begin{document}

\section{Introduction}

Stars more massive than $\geq$20\,$M_{\odot}$ experience the
short-lived luminous blue variable (LBV) stage \cite{Conti1984},
which among other evolutionary stages of massive stars is the most
interesting in observational manifestations and perhaps the most
important in the evolutionary sense
\cite{HD1994,L1994,vG2001,Groh2013}. During this stage a massive
star exhibits irregular spectroscopic and photometric variability
on time-scales from years to decades or longer, which is reflected
in changes of the stellar type from late O/early B supergiants to
A/F-type ones (see e.g. \cite{Stahl2001,Groh2009}) and changes in
the brightness by several magnitudes. At the brightness maximum,
LBVs could be confused with supernovae (e.g.
\cite{Good1989,VDyk2002}), and it is believed that some LBVs could
be the direct progenitors of supernovae (e.g.
\cite{KV2006,Groh2013}). The LBV stars experience episodes of
enhanced, sometimes eruptive, mass loss, so that most of them (see
\cite{Clark2005,KGB2015a} and Table\,\ref{LBV}) are surrounded by
compact ($\sim 0.1-1$ pc in diameter) shells with a wide range of
morphologies (e.g. \cite{Nota1995,Weis2001,GK2010c}).

The LBV phenomenon is still ill-understood, which is mostly
because the LBV stars are very rare objects. The recent census of
Galactic confirmed and candidate LBVs (cLBVs) presented in
\cite{Vink2012} lists only 13 and 25 stars, respectively. The
discovery of additional LBVs would, therefore, be of great
importance for understanding their evolutionary status and their
connection to other massive transient stars, as well as for
unveiling the driving mechanism(s) of the LBV phenomenon.

Detection of LBV-like shells can be considered as an indication
that their associated stars are massive and evolved, and therefore
could be used for the selection of candidate massive stars for
follow-up spectroscopy. Because of the huge interstellar
extinction in the Galactic plane, the most effective channel for
the detection of circumstellar shells is through imaging with
modern infrared (IR) telescopes. Application of this approach
using the {\it Spitzer Space Telescope} and {\it Wide-field
Infrared Survey Explorer} ({\it WISE}) resulted in the discovery of
hundreds of such shells whose central stars could be LBVs or other
types of evolved massive stars
(\cite{GK2010c,Wach2010,Miz2010,GK2011}). Indeed, follow-up
optical and IR spectroscopy of these central stars led to the
discovery of dozens of new cLBV, blue supergiant and Wolf-Rayet
(WR) stars in the Milky Way
(\cite{G2009,GK2010a,GK2010b,GK2010c,Wach2010,Wach2011,GK2012,Str2012a,Str2012b,Burgm2013,Flag2014,
GM2014,GC2014,GK2014,KGB2015a,GK2015a,GK2015b,KGB2015b}). Because
of reddening many of the central stars are very dim in the optical,
which makes inevitable the use of 8--10-m class telescopes like
the Southern African Large Telescope (SALT). Here we report the
results of optical spectroscopy with the SALT of 54 central stars
of compact mid-IR nebulae discovered with {\it Spitzer} and {\it
WISE}.

%%%%%%%%%%%%%%%%%%%%%%%%%%%%%%%%%%%%%%%%%%%%%%%%%%%%%%%%%%%%%%%%%%%%%%%%%%%
\begin{figure*}
    \begin{center}
    \includegraphics[angle=0,width=14cm,height=16.0cm,clip=]{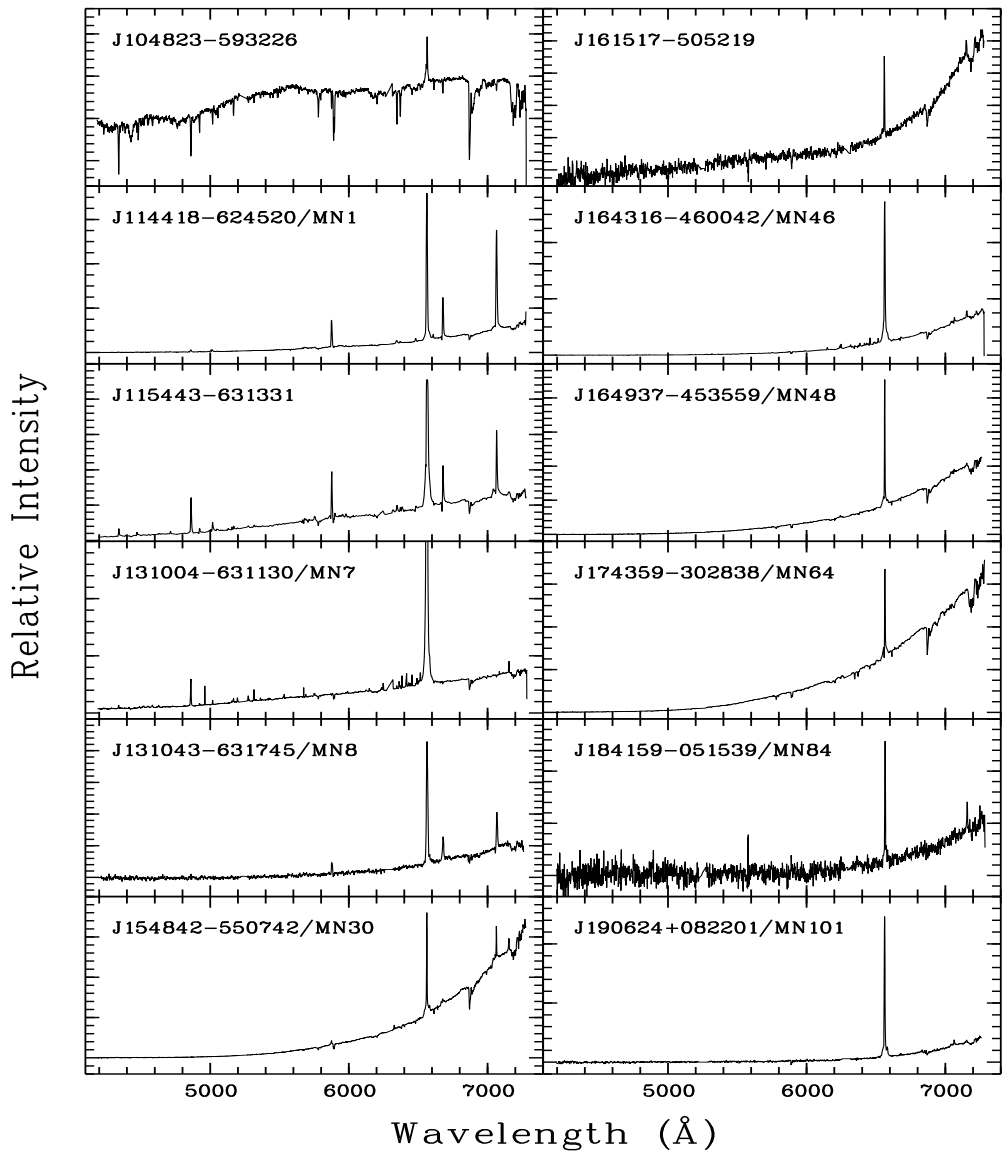}
    \caption{The observed SALT spectra of a dozen emission-line stars from
    our sample of candidate evolved massive stars revealed with {\it Spitzer}
    and {\it WISE}.}
    \label{spec}
    \end{center}
\end{figure*}
%%%%%%%%%%%%%%%%%%%%%%%%%%%%%%%%%%%%%%%%%%%%%%%%%%%%%%%%%%%%%%%%%%%%%%%%%%%

\section{Observations}
\label{obs}

\begin{figure*}
    \begin{center}
    \includegraphics[angle=0,width=13.0cm,clip=]{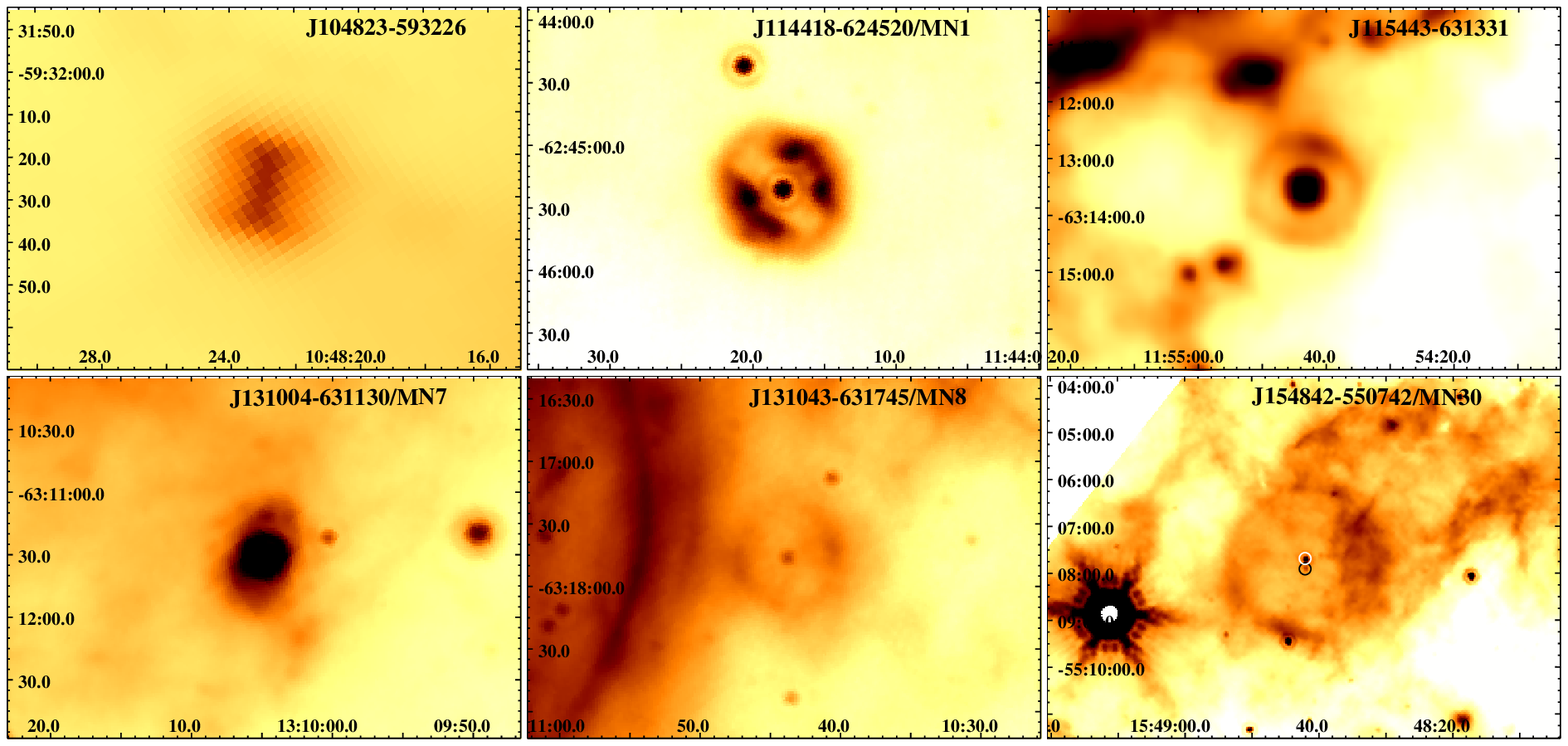}
    \includegraphics[angle=0,width=13.0cm,clip=]{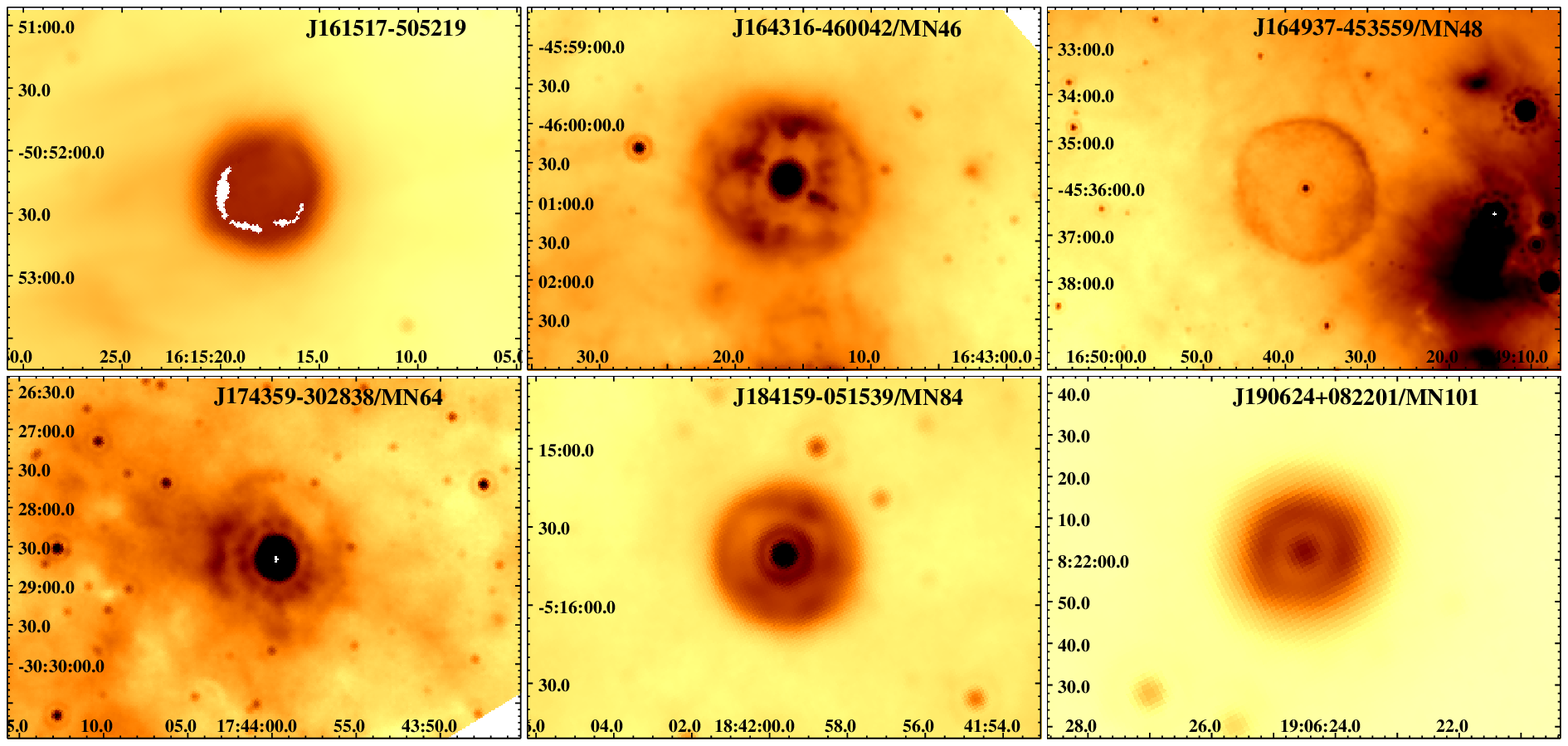}
    \caption{Mid-IR images of circumstellar nebulae around stars whose
    spectra are shown in Figure\,1. All but one of these nebulae
    were discovered with {\it Spitzer} at 24\,$\mu$m. The nebula around
    J115443-631331 was discovered with {\it WISE} at 22\,$\mu$m. The nebula
    shown in the right panel of the second row contains two stars near the
    geometrical centre. One of them (marked by a white circle) is the cLBV
    J154842-550742/MN30. The second one (marked by a black circle) is
    the WC9 star J154842-550755. The coordinates are in units of RA(J2000) and
    Dec.(J2000) on the horizontal and vertical scales,
    respectively.}
    \label{neb}
    \end{center}
\end{figure*}

The SALT observations were carried out in 2010--2015 with the
Robert Stobie Spectrograph (RSS; \cite{Burg2013}) in the long-slit
mode. In most cases, the spectra covered the range of 4200$-$7300
\AA. The primary reduction of the data was done with SALT
science pipeline. After that, the bias and gain corrected and
mosaicked long-slit data were reduced in the way described in
\cite{Kniazev2008}. Examples of 1D finally reduced spectra of a
dozen emission-line stars are shown in Figure~\ref{spec}, while
Figure\,\ref{neb} presents the mid-IR images of circumstellar
nebulae associated with these stars. The list of all observed
targets is given in Table\,\ref{list}.

%%%%%%%%%%%%%%%%%%%%%%%%%%%%%%%%%%%%%%%%%%%%%%%%%%%%%%%%%%%%%%%%%%%%%%%%%%%
%%%%%%%%%%%%%%%%%%%%%%%%%%%%%%%%%%%%%%%%%%%%%%%%%%%%%%%%%%%%%%%%%%%%%%%%%%%

To search for possible spectroscopic and photometric variability
of the newly identified cLBVs, we obtained additional spectra with
the SALT and performed photometric monitoring of these stars with
the 76-cm telescope of the South African Astronomical Observatory.

\section{Results}
\label{res}

\begin{figure*}
    \begin{center}
    \includegraphics[angle=270,width=14.0cm,clip=]{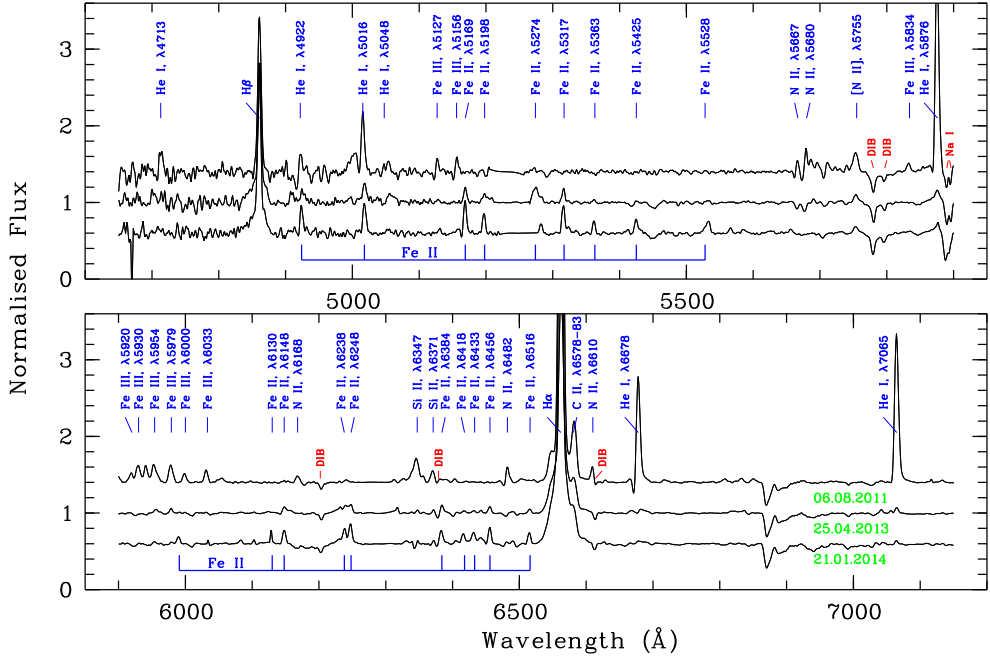}
    \caption{Evolution of the (normalized) spectrum of the new Galactic
    bona fide LBV Wray 16-137 in 2011--2014 (adopted from \cite{GK2014}).}
    \label{Wray}
    \end{center}
\end{figure*}

We carried out optical SALT spectroscopy of 54 candidate evolved
massive stars. The first results of our observing program were
presented in
\cite{GK2012,Todt2013,GC2014,GK2014,GK2015a,KGB2015a,GK2015b}.
Table\,\ref{list} summarizes the (mostly preliminary) spectral
classification of the observed targets. We detected about two
dozen of emission-line stars, of which 15 stars were classified
as cLBVs. Subsequent spectroscopic and photometric monitoring of
these stars allowed us to confirm the LBV status of four of them
(see Tables\,\ref{list} and
\cite{GK2014,KGB2015a,GK2015b,KGB2015b}). Figure\,\ref{Wray} shows
the evolution of the spectrum of Wray\,16-137 (one of the four newly
identified Galactic bona fide LBVs) in 2011--2014. One can see
that the He\,{\sc i} emission lines have almost disappeared, while
numerous Fe\,{\sc ii} emissions have become prominent. These changes
along with significant brightness increase of the star (by about 1
mag during three years) indicate that currently Wray\,16-137
experiences an S\,Dor-like outburst.

%%%%%%%%%%%%%%%%%%%%%%%%%%%%%%%%%%%%%%%%%%%%%%%%%%%%%%%%%%%%%%%%%%%%%%%%%%%
\begin{table*}
\begin{center}
%\begin{minipage}{\textwidth}
\caption{The observed central stars of mid-IR nebulae and their
(mostly preliminary) spectral classification. The stars whose
spectra and nebulae are presented in Figures\,1 and 2 are
starred.} \label{list} \small
\begin{tabular}{lll||lll} \\ \hline %\hline
\MC{1}{c}{ Object }      & \MC{1}{c}{ Name   }      & \MC{1}{c||}{
Type  }      & \MC{1}{c}{ Object }      & \MC{1}{c}{ Name   } &
\MC{1}{c}{ Type   }      \\

\MC{1}{c}{ (1) } & \MC{1}{c}{ (2) } & \MC{1}{c||}{ (3) } &
\MC{1}{c}{ (1) } & \MC{1}{c}{ (2) } &
\MC{1}{c}{ (3) } \\
\hline
%\[-0.3cm]
 J045304-692352 & BAT99\,3a        & WN3b+O6\,V$^a$  & J153856-563722        & MN25         & OB           \\
 J052848-705105 &                  & OB              & J154527-535602        & MN26         & OB           \\
 J052412-683011 & LHA\,120\,N      & OB              & J154842-550742$^\ast$ & MN30         & cLBV$^i$     \\
 J071810-265124 & CD$-$26\,4148    & OB              & J154842-550755        &              & WC9$^j$      \\
 J091714-495502 &                  & M               & J161132-512906        & MN40         & OB           \\
 J100122-550046 &                  & OB              & J161517-505219$^\ast$ &              & cLBV$^k$     \\
 J103638-580012 &                  & OB              & J163239-494213        & MN44         & LBV$^l$      \\
 J104823-593226$^\ast$ & HD\,93795 & A               & J164316-460042$^\ast$ & MN46         & cLBV$^m$     \\
 J110340-592559 & HD\,96042        & OB              & J164937-453559$^\ast$ & MN48         & LBV$^n$      \\
 J111428-611820 & Wray\,15-780     & OB              & J170723-395651        & MN50         & M$^o$        \\
 J114418-624520$^\ast$ & MN1       & cLBV$^b$        & J171307-384734        & CD$-$38\,11646 & OB         \\
 J115443-631331$^\ast$ &           & cLBV            & J172031-330949        & WS2          & cLBV$^p$     \\
 J120058-631259 & MN2              & OB              & J173753-302311        &              & OB           \\
 J124626-632427 & WRAY\,17-56      & PN$^c$          & J173918-312424        & MN59         & OB$^q$       \\
 J131004-631130$^\ast$ & MN7       & cLBV            & J174359-302838$^\ast$ & MN64         & cLBV$^r$  \\
 J131028-621331 &                  & OB              & J174627-302001        &              & OB           \\
 J131043-631745$^\ast$ & MN8       & cLBV$^d$        & J180433-210326        & HD\,313642   & A            \\
 J131933-623844 & MN10             & OB$^e$          & J180612-211745        & MN74         & OB           \\
 J132647-615924 &                  & M               & J180823-221939        & ALS\,4684    & OB           \\
 J133628-634538 & WS1              & LBV$^f$         & J182721-132209        &              & OB           \\
 J133654-632552 &                  & OB              & J183217-091614        &              & OB$^s$       \\
 J135015-614855 & Wray\,16-137     & LBV$^g$         & J183528-064415        &              & OB           \\
 J140707-652934 & CPD$-$64\,2731   & O               & J184159-051539$^\ast$ & MN84         & cLBV$^t$     \\
 J143111-610202 & MN14             & OB              & J184246-031317        &              & [WN5]$^u$    \\
 J151342-585318 & MN17             & OB              & J185404+033544        &              & M            \\ %WS4
 J151641-582226 & MN18             & B1\,Ia$^h$      & J190421+060001        & MN100        & OB$^v$       \\
 J151959-572415 & MN19             & OB              & J190624+082201$^\ast$ & MN101        & cLBV$^w$     \\
%\hline
\hline \MC{6}{p{15cm}}{{\it Notes}: $^a$ \cite{GC2014}; $^b$
classified as an Oe/WN in \cite{Wach2010}; $^c$ originally
classified as a PN in \cite{Par2006}; $^d$ classified as an Oe/WN
in \cite{Wach2010}; $^e$ originally classified as an OB in
\cite{Wach2011}; $^f$ \cite{GK2012,KGB2015a}; $^g$ \cite{GK2014};
$^h$ \cite{GK2015a}; $^i$ classified as a Be/B[e]/LBV in
\cite{Wach2011}; $^j$ originally classified as a WC9 in
\cite{Wach2011}; $^k$ classified as a star in transition from AGB
to PN in \cite{Van1989}; $^l$ \cite{GK2015b}; $^m$ \cite{GK2010c};
$^{n}$ \cite{KGB2015b}; $^o$ classified as a M1\,I in
\cite{Wach2010}; $^p$ \cite{GK2012}; $^q$ originally classified as
an OB in \cite{Wach2010}; $^r$ classified as a Be in
\cite{Wach2010} and as an OB in \cite{Wach2011}; $^s$ classified
as a BA in \cite{Wach2011}; $^t$ classified as a Be/B[e]/LBV in
\cite{Wach2010} and as a cLBV in \cite{Str2012b}; $^u$ originally
classified as a WN6 in \cite{Wach2010} and then re-classified as a
[WN5] in \cite{Todt2013}; $^v$ classified as a FG in
\cite{Wach2011}; $^w$ classified as B[e]/LBV in \cite{Wach2011}
and as a cLBV in \cite{Str2012b}.}
\end{tabular}
\end{center}
\end{table*}
%%%%%%%%%%%%%%%%%%%%%%%%%%%%%%%%%%%%%%%%%%%%%%%%%%%%%%%%%%%%%%%%%%%%%%%%%%%

%%%%%%%%%%%%%%%%%%%%%%%%%%%%%%%%%%%%%%%%%%%%%%%%%%%%%%%%%%%%%%%%%%%%%%%%%%%
%%%%%%%%%%%%%%%%%%%%%%%%%%%%%%%%%%%%%%%%%%%%%%%%%%%%%%%%%%%%%%%%%%%%%%%%%%%

We also discovered a rare WN-type central star of a planetary
nebula (Abell\,48), which is the second known example of [WN]
stars \cite{Todt2013}. Thanks to the high angular resolution of
{\it Spitzer} images (6 arcsec at 24\,$\mu$m), we detected a new
circular shell in the Large Magellanic Cloud. Follow up
spectroscopy of its central star with SALT (and several other
telescopes) resulted in the discovery of a new WR star in a close,
eccentric binary system with an O6\,V star \cite{GC2014}.
The majority of the remaining targets were tentatively classified
as OB, A and M stars.

%\section{Galactic bona fide LBVs}
%\label{sensus}

Finally, we present in Table\,\ref{LBV} the current census of the
Galactic bona fide LBVs (eighteen stars in total). The objects
with detected circumstellar nebulae are starred. As follows from
the table, 72 per cent of the Galactic confirmed LBVs are
associated with nebulae. This provides further proof that
the detection of compact mid-IR shells is a powerful tool for
identifying (candidate) LBVs. Searches for new mid-IR nebulae
continue, with further discoveries of LBVs and other related stars
anticipated.

%==========================================================================
\begin{table*}
\caption{Current census of the Galactic bona fide LBVs. The
objects with detected circumstellar nebulae are starred.}
\label{LBV}
\begin{tabular}{p{2.8cm}p{2.4cm}p{3.0cm}p{5.2cm}}
\hline HR\,Car$^\ast$ & $\eta$\,Car$^\ast$ & AG\,Car$^\ast$ & Wray\,15-751$^\ast$ \\
$[$GKM2012$]$\,WS1$^\ast$ & Wray\,16-137$^\ast$ & $[$GKF2010$]$\,MN44$^\ast$ &
%$[$GKF2010$]$\,MN46$^\ast$ \\
Cl*\,Westerlund\,1\,W\,243 \\
$[$GKF2010$]$\,MN48$^\ast$ & HD\,160529 & GCIRS\,34W & $[$MMC2010$]$\,LBV\,G0.120$-$0.048$^\ast$ \\
qF\,362 & HD\,168607 & MWC\,930$^\ast$ & G24.73+0.69$^\ast$ \\
AFGL\,2298$^\ast$ & P\,Cyg$^\ast$ & & \\
\hline
\end{tabular}
\end{table*}
%==========================================================================

%\section{Acknowledgements}
\acknowledgments{
The observations reported in this paper were obtained with the SALT
programs \mbox{2010-1-RSA\_OTH-001}, \mbox{2011-3-RSA\_OTH-002},
\mbox{2013-1-RSA\_OTH-014}, \mbox{2013-2-RSA\_OTH-003} and
\mbox{2015-1-SCI-017}. AYK acknowledges support from the National
Research Foundation (NRF) of South Africa. VVG acknowledges the
Russian Science Foundation grant 14-12-01096.
This work was partially supported by the Russian Foundation
for Basic Research grant 16-02-00148.
We are grateful to referees for useful suggestions on the manuscript.
}


\begin{thebibliography}{99}

\bibitem{Burg2013} E.B. Burgh et al., \emph{Prime focus imaging spectrograph for the Southern African Large Telescope:
optical design, SPIE} {\bf 4841} (2003) 1463

\bibitem{Burgm2013} S. Burgmeister et al. \emph{WR 120bb and WR 120bc: a pair of WN9h stars with
            possibly interacting circumstellar shells, MNRAS} {\bf 429} (2013) 3305

\bibitem{Clark2005} J.S. Clark, V.M. Larionov, A. Arkharov,
           \emph{On the population of galactic Luminous Blue Variables, A\&A}
           {\bf 435} (2005) 239

\bibitem{Conti1984} P.S. Conti, \emph{Basic observational constraints on the evolution of massive stars},
            in proceedings of \emph{Observational Tests of the Stellar Evolution Theory} (1984) 233

\bibitem{Flag2014} N. Flagey et al., \emph{Palomar/TripleSpec observations of Spitzer/MIPSGAL 24 $\mu$m circumstellar shells:
unveiling the natures of their central sources, AJ}
           {\bf 148} (2014) 34

\bibitem{Good1989} R. W. Goodrich et al., \emph{SN 1961V - an extragalactic Eta Carinae analog, ApJ} {\bf 342} (1989) 908

\bibitem{Groh2009} J.H. Groh et al., \emph{Bona fide, strong-variable Galactic luminous blue variable
           stars are fast rotators: detection of a high rotational velocity in HR Carinae, ApJ} {\bf 705} (2009) L25

\bibitem{Groh2013} J.H. Groh et al., \emph{Massive star evolution: luminous blue variables as unexpected supernova progenitors, A\&A} {\bf 550} (2013) L7

\bibitem{G2009}   V.V. Gvaramadze et al., \emph{Discovery of a new Wolf-Rayet star and its ring nebula in Cygnus, MNRAS} {\bf 400} (2009) 524

\bibitem{GK2010a} V.V. Gvaramadze et al., \emph{A new Wolf-Rayet star and its circumstellar nebula in Aquila, MNRAS} {\bf 403} (2010a) 760

\bibitem{GK2010b} V.V. Gvaramadze et al., \emph{MN112: a new Galactic candidate luminous blue variable, MNRAS} {\bf 405} (2010b) 520

\bibitem{GK2010c} V.V. Gvaramadze, A.Y. Kniazev, S. Fabrika, \emph{Revealing evolved massive stars with Spitzer, MNRAS} {\bf 405} (2010c) 1047

\bibitem{GK2011}  V.V. Gvaramadze et al., \emph{Search for OB stars running away from young star clusters. II. The NGC 6357 star-forming region, A\&A} {\bf 535} (2011) A29

\bibitem{GK2012}  V.V. Gvaramadze et al., \emph{Discovery of two new Galactic candidate luminous blue variables with Wide-field Infrared Survey Explorer, MNRAS} {\bf 421} (2012) 3325

\bibitem{GM2014}  V.V. Gvaramadze et al., \emph{TYC 3159-6-1: a runaway blue supergiant, MNRAS} {\bf 437} (2014) 2761

\bibitem{GC2014}  V.V. Gvaramadze et al., \emph{Discovery of a new Wolf-Rayet star and a candidate star cluster in the Large Magellanic Cloud with Spitzer, MNRAS} {\bf 442} (2014) 929

\bibitem{GK2014}  V.V. Gvaramadze et al., \emph{Discovery of a new Galactic bona fide luminous blue variable with Spitzer, MNRAS} {\bf 445} (2014) L84

\bibitem{GK2015a} V.V. Gvaramadze et al., \emph{The blue supergiant MN18 and its bipolar circumstellar nebula, MNRAS} {\bf 454} (2015) 219

\bibitem{GK2015b} V.V. Gvaramadze, A.Y. Kniazev, L. Berdnikov, \emph{Discovery of a new bona fide luminous blue variable in Norma, MNRAS} {\bf 454} (2015) 3710

\bibitem{HD1994} R.M. Humphreys, K. Davidson, \emph{The luminous blue variables: astrophysical geysers, PASP} {\bf 106} (1994) 1025

\bibitem{Kniazev2008} A.Y. Kniazev et al., \emph{The metallicity extremes of the Sagittarius dSph: SALT spectroscopy of PNe, MNRAS} {\bf 388} (2008) 1667

\bibitem{KGB2015a} A.Y. Kniazev, V.V. Gvaramadze, L.N. Berdnikov,
\emph{WS1: one more new Galactic bona fide luminous blue variable, MNRAS} {\bf 449} (2015) L60

\bibitem{KGB2015b} A.Y. Kniazev,  V.V. Gvaramadze, L.N. Berdnikov, \emph{in preparation} (2016)

\bibitem{KV2006} R. Kotak, J.S. Vink, \emph{Luminous blue variables as the progenitors of supernovae with quasi-periodic radio modulations, A\&A} {\bf 460} (2006) L5

\bibitem{L1994} N. Langer et al., \emph{Towards an understanding of very massive stars. A new evolutionary scenario relating O stars, LBVs and Wolf-Rayet stars, A\&A} {\bf 290} (1994) 819

\bibitem{Miz2010} D.R. Mizuno et al., \emph{A catalog of MIPSGAL disk and ring sources, AJ} {\bf 139} (2010) 1542

\bibitem{Nota1995} A. Nota et al., \emph{Nebulae around luminous blue variables: a unified picture, ApJ} {\bf 448} (1995) 788

\bibitem{Par2006} Q.A. Parker et al., \emph{The Macquarie/AAO/Strasbourg H$\alpha$ planetary nebula catalogue: MASH, MNRAS} {\bf 373} (2006) 79

\bibitem{Stahl2001} O. Stahl et al., \emph{Long-term spectroscopic monitoring of the luminous blue variable AG Carinae, A\&A} {\bf 375} (2001) 54

\bibitem{Str2012a} G.S. Stringfellow et al., \emph{New  Galactic candidate luminous blue variables and Wolf-Rayet stars},
in proceedings of \emph{From Interacting Binaries to Exoplanets:
Essential Modeling Tools, IAUS} {\bf 282} (2012) 267

\bibitem{Str2012b} G.S. Stringfellow et al., \emph{Spectral identification of new Galactic cLBV and WR Stars},
in proceedings of \emph{Scientific Meeting in Honor of Anthony F.
J. Moffat held at Auberge du Lac Taureau, ASPC} {\bf 465}(2012) 514

\bibitem{Todt2013} H. Todt et al., \emph{Abell 48 - a rare WN-type central star of a planetary nebula, MNRAS} {\bf 430} (2013) 2302

\bibitem{Van1989} W.E.C.J. van der Veen, H.J. Habing, T.R. Geballe,
          \emph{Objects in transition from the AGB to the planetary nebula stage - New visual and infrared observations, A\&A} {\bf 226} (1989) 108

\bibitem{vG2001} A.M. van Genderen, \emph{S Doradus variables in the Galaxy and the Magellanic Clouds, A\&A} {\bf 366} (2001) 508

\bibitem{VDyk2002} S.D. Van Dyk et al., \emph{Possible recovery of SN 1961V in Hubble Space Telescope archival images, PASP} {\bf 114} (2002) 700

\bibitem{Vink2012} J.S. Vink, \emph{Eta Carinae and the luminous blue variables} in proccedings of \emph{Eta Carinae and the Supernova Impostors, Astrophys. \& Sp. Sci. Library} {\bf 384}
% (ed. K. Davidson \& R.M. Humphreys, Springer Media, New York)
(2012) 221

\bibitem{Wach2010} S. Wachter et al., \emph{A hidden population of massive stars with circumstellar shells discovered with the Spitzer Space Telescope, AJ} {\bf 139} (2010) 2330

\bibitem{Wach2011} S. Wachter et al. \emph{Massive stars with circumstellar shells discovered with the \textit{Spitzer} Space Telescope, BSRSL} {\bf 80} (2011) 322

\bibitem{Weis2001} K. Weis, \emph{LBV nebulae: The mass lost from the most massive stars, RvMA} {\bf 14} (2001) 261

\end{thebibliography}
\end{document}